\documentclass[12pt,preprint]{aastex}

\def\Mfo   {M~51}
\def\Msf   {M~74}
\def\Nste  {NGC~628}
\def\Nottt {NGC~1232}
\def\Ntoef {NGC~3184}
\def\Nfonf {NGC~5194}

\def\m    {\phantom{-}}
\def\o    {\phantom{1}}
\def\p    {\phantom{0}}

\def\deg  {\ifmmode {^\circ}\else {$^\circ$}\fi}
\def\porm {\ifmmode {\pm}\else {$\pm$}\fi}
\def\chisqpdf {\ifmmode {\chi^2_{\rm pdf}}\else {$\chi^2_{\rm pdf}$}\fi}
\def\chisq    {\ifmmode {\chi^2}\else {$\chi^2$}\fi}

\def\Rref {\ifmmode {R_{{\rm ref}}}\else {$R_{{\rm ref}}$}\fi}
\def\b    {\ifmmode {\beta}\else {$\beta$}\fi}
\def\bref {\ifmmode {\beta_{{\rm ref}}}\else {$\beta_{{\rm ref}}$}\fi}

\def\etal {et al.~}
\def\eg   {e.g.,~}

\def\Ha   {H$\alpha$}
\def\HII   {H~{\footnotesize II}}

\def\d    {\ifmmode {{\rlap{.}}^\circ}\else {${\rlap{.}}^\circ$}\fi}
\def\s    {\ifmmode {{\rlap{.}}^s}\else {${\rlap{.}}^s$}\fi}
\def\as   {\ifmmode {{\rlap{.}}^{''}}\else {${\rlap{.}}^{''}$}\fi}

\newbox\grsign \setbox\grsign=\hbox{$>$} \newdimen\grdimen \grdimen=\ht\grsign
\newbox\laxbox \newbox\gaxbox
\setbox\gaxbox=\hbox{\raise.5ex\hbox{$>$}\llap
     {\lower.5ex\hbox{$\sim$}}}\ht1=\grdimen\dp1=0pt
\setbox\laxbox=\hbox{\raise.5ex\hbox{$<$}\llap
     {\lower.5ex\hbox{$\sim$}}}\ht2=\grdimen\dp2=0pt
\def\gax{\mathrel{\copy\gaxbox}}
\def\lax{\mathrel{\copy\laxbox}}
\shorttitle{Spiral Arms} 
\shortauthors{Honig \& Reid}

\begin{document}

\title{Characteristics of Spiral Arms in Late-type Galaxies}

\author{Z. N. Honig\altaffilmark{1} \& M. J. Reid\altaffilmark{1}}

\altaffiltext{1}{Harvard-Smithsonian Center for
   Astrophysics, 60 Garden Street, Cambridge, MA 02138, USA}

\begin{abstract}
We have measured the positions of large numbers of \HII\ regions in four 
nearly face-on, late-type, spiral galaxies:
\Nste\ (\Msf), \Nottt, \Ntoef\ and \Nfonf\ (\Mfo).   Fitting log-periodic 
spiral models to segments of each arm yields local estimates of spiral pitch 
angle and arm width.  While pitch angles 
vary considerably along individual arms, among arms within a galaxy, and among 
galaxies, we find no systematic trend with galactocentric distance.  
We estimate the widths of the arm segments from the scatter in the distances 
of the \HII\ regions from the spiral model.   All major arms in these galaxies 
show spiral arm width increasing with distance from the galactic center, 
similar to the trend seen in the Milky Way.  However, in the outer-most parts 
of the galaxies, where massive star formation declines, some arms reverse this 
trend and narrow.  We find that spiral arms often appear to be composed
of segments of $\sim5$ kpc length, which join to form kinks and abrupt changes 
in pitch angle and arm width; these characteristics are consistent with 
properties seen in the large $N$-body simulations of \citet{DOnghia:13} and others.

\end{abstract}

\keywords{galaxies: individual (\Mfo, \Msf, \Nottt, \Ntoef) --
          galaxies: spiral -- galaxies: structure -- \HII\ regions}

\section{Introduction}

Spiral structures in galaxies have been extensively studied, including
the nature and number of arms and values for pitch angles, 
which measure how tightly wound are the spirals.   One characteristic
of spirals that has received little attention is the width of arms.
For the Milky Way, recent measurements of parallax of molecular masers 
associated with newly formed massive stars have indicated that the widths
of spiral arms, estimated from the scatter of these sources about a segment
of a spiral arm, increases with Galactocentric distance \citep{Reid:14}.

Assuming that the Milky Way is not atypical among spiral galaxies, 
one might expect a similar increase in arm widths with radius in 
other late-type spirals.  Whereas the measurement of arm width for the
Milky Way is difficult, owing to our location within the Galactic disk
and the difficulty of obtaining accurate distance measurements at characteristic 
distances from the Sun of $\sim5$ kpc, it should be straight-forward in external
galaxies.   Thus, we undertook a detailed study of the patterns of giant 
\HII\ regions in four nearby, nearly face-on, late-type galaxies 
that display clear spiral structure.  

We use the locations of giant \HII\ regions as tracers of spiral structure.
The galaxy images used for measurement of the locations of 
\HII\ regions are documented in Section \ref{sect:measurements}, along
with the methods used to determine positions.  Our approach to fitting 
segments of spiral patterns to these data is presented in Section \ref{sect:fitting}.   
The results for each galaxy are given in Section \ref{sect:results}.   
Finally, in Section \ref{sect:characteristics}, we discuss
the characteristics of the spiral arms and compare these characteristics to
those of the Milky Way.

\section{Measured Positions of \HII\ Regions} \label{sect:measurements}

Table~\ref{table:galaxies} lists information about the images of the
four galaxies studied in this paper.  All of the images are publicly 
available on the internet, and we selected high resolution images in \Ha\ 
or blue-filtered emission to maximize the contrast of \HII\ regions with
respect to the background continuum emission from the galaxy.  
Images were either downloaded in the Flexible Image Transport System 
(FITS) format or were converted to that format using the ImageMagick
program  ``convert'' available at {\url {http://www.imagemagick.org}}.
We also used this software to prepare images for display in our figures.

\begin{deluxetable}{lllcclcc}
\tablecolumns{8} \tablewidth{0pc}
\tablecaption{}
\tablehead {
  \colhead{Galaxy} & \colhead{Alias}& \colhead{Type} & \colhead{Distance} &
  \colhead{Telescope} & \colhead{Filter} &
  \colhead{Pixel Scale}  & \colhead{Reference/} 
\\
  \colhead{}      &\colhead{}      & \colhead{} & \colhead{(Mpc)} &
  \colhead{} & \colhead{} & 
  \colhead{(arcsec)}& \colhead{Source}  
          }
\startdata
\Nste    &\Msf   &SA(s)c     &10.0   &HST   &\Ha  &0.150  &1 \\
\Nottt   &       &SAB(rs)c   &21.0   &VLT   &Blue &0.041  &2 \\
\Ntoef   &       &SAB(rs)cd  &\p8.2  &JKT   &\Ha  &0.243  &3 \\ 
\Nfonf   &\Mfo   &SA(s)bc    &\p9.4  &HST   &\Ha  &0.050  &4 \\
\enddata
\tablecomments{\footnotesize
Assumed distances used to convert pixel to linear scales. References:
1: {\url {http://heritage.stsci.edu/2007/41/fast\_facts.html}};
2: {\url {https://www.eso.org/public/usa/images/eso9845d}};
3: {\url {http://cdsarc.u-strasbg.fr/viz-bin/qcat?J/A+A/426/1135}},
\citet{Knapen:04};
4: {\url {http://archive.stsci.edu/prepds/m51/datalist.html}}.
Telescopes: HST: Hubble Space Telescope; VLT: Very Large Telescope; 
JKT: Jacobus Kapteyn Telescope.
                }
\label{table:galaxies}
\end{deluxetable}

FITS format images were loaded into the National Radio Astronomy Observatory's
\footnote{The National Radio Astronomy Observatory is a facility of the National
Science Foundation operated under cooperative agreement by Associated Universities, Inc.}
Astronomical Image Processing System (AIPS).  We used the task JMFIT to estimate
positions by fitting two-dimensional elliptical Gaussian brightness distributions 
to individual sub-regions within an image that each contained one \HII\ region. 
Foreground stars in the Milky Way were easily recognized as point-like objects, 
usually with diffraction spikes, and excluded.  Contamination of our \HII\
region sample by other objects (\eg supernova remnants or planetary nebulae) 
should not be significant for these \Ha\ or blue-filtered images. 
 
Fitted positions were obtained in pixel units, and we used the pixel scale and 
assumed galaxy distance to translate these to physical scales in kiloparsec units.   
Note that we often adjusted the image display
transfer functions when preparing sub-regions of a galaxy for measurement, and that
many \HII\ regions are not visible on the low dynamic-range figures published in 
this paper.  We also measured the position of the bright center of each galaxy and 
subtracted this position from each \HII\ region position in order to translate to 
galactocentric coordinates. We made no attempt to deproject the \HII\ region
positions for these galaxies, which are within about $30^\circ$ from a face-on 
orientation.
The galactic positions for all \HII\ regions are available as on-line material.  

\section {Fitting Spiral Arm Segments} \label{sect:fitting}

We fit segments of spiral arms to the positions of \HII\ regions  
using a Bayesian Markov chain Monte Carlo technique.  
The adopted model is a log-periodic spiral given by
$$R = \Rref~e^{-(\b-\bref)\tan{\psi}}~~~,\eqno(1)$$
where \b\ is galactocentric azimuth defined as zero toward the north and increasing
east of north, \Rref\ is the galactocentric radius at azimuth \bref, and $\psi$ is
the spiral pitch angle.   The parameter \bref\ was assigned the average of the
measured \HII\ region azimuths, leaving two parameters to be solved for: \Rref\ and $\psi$.

Since, the precision in measuring an \HII\ region position ($\lax30$ pc) is
significantly better than the intrinsic scatter in their positions about a best-fit spiral 
segment ($\gax100$ pc), we also solved for an arm-width parameter ($\sigma_w$) to characterize
the scatter and to allow for this source of astrophysical noise when fitting a spiral.  
(While there can be many definitions of arm width, throughout this paper we adopt a 
Gaussian ($1\sigma$) approximation to the distribution of the minimum distances
of \HII\ regions from the model arm segment.)   We calculated likelihood functions from
the weighted residuals, $r_i = \Delta d_i / \sigma_w$, where $\Delta d_i$ is 
the minimum distance of the $i^{th}$ \HII\ region from the model spiral (evaluated numerically)
and $\sigma_w$ is its ``uncertainty'' owing to intrinsic scatter within the arm.

We randomly generated trials (Markov chains) of the two parameters characterizing 
the spiral arm segment (\Rref\ and $\psi$) and the parameter characterizing the 
width of the spiral arm ($\sigma_w$) and accepted/rejected these trials with
the Metropolis-Hastings algorithm.  
We employed two likelihood functions in our analysis: an ``error-tolerant'' 
likelihood function \citep[see ``a conservative formulation'' by][]{Sivia:06}
given by 
$$\prod_{i=1}^N ~ {1\over\sigma_w}~{ {1 - e^{-r_i^2/2}} \over r_i^2 }~~\eqno(2)$$
and a least-squares likelihood function given by
$$-\prod_{i=1}^N ~ {1\over\sigma_w}~e^{-r_i^2/2}~~,\eqno(3)$$  
where $N$ is the number of measured \HII\ regions.
Note the factor ${1\over\sigma_w}$ in the likelihoods, often a constant
and ignored, is retained here since we vary its value in order to estimate arm width.

The error-tolerant likelihood function is relatively insensitive to outlying data 
points and was used to find and then remove interarm \HII\ regions from the sample
in an unbiased fashion.  
We accomplished this by fitting with the Eq. (2) likelihood and discarding \HII\ 
regions with $>3\sigma$ deviations.   After this preliminary editing, we fitted the 
spiral with the least-squares likelihood (Eq. 3).  Usually, at this stage,
all residuals were $<3\sigma$, but occasionally one or two \HII\ regions with
residuals just above $3\sigma$ were encountered.  In these cases, the outliers were 
removed from the sample and the spiral re-fitted.

We first fitted a single log-periodic spiral to all \HII\ regions in a given arm.
Such ``global'' fits usually revealed systematic departures from a constant pitch 
angle model.  Since we would like to investigate the possible increase in arm width 
along the arm, we re-fit each arm with spiral segments of length roughly 5 to 10 kpc.   
The boundaries for the arm segments were chosen based on three criteria: 
1) breaks in the density or scatter of \HII\ regions,
2) apparent changes in the local pitch angles among segments, and
3) to keep comparably sized samples of \HII\ regions among the segments in one arm.
Fitting short arm segments also has the advantage of minimizing systematic error in the
width estimates, owing to the simplification of the assumed spiral model.

Marginalized posteriori probability density functions (PDFs) were obtained
from binned histograms of the individual parameter values for all trials. 
Since the PDFs were approximately Gaussian in shape, we
adopted the center and half-width of the 68\% confidence intervals as best
estimates of the parameter values and $1\sigma$ uncertainties.

\section {Results} \label{sect:results}

\subsection{\Nste\ (\Msf)}

\begin{figure}[!htbp]
\epsscale{0.85} 
\plotone{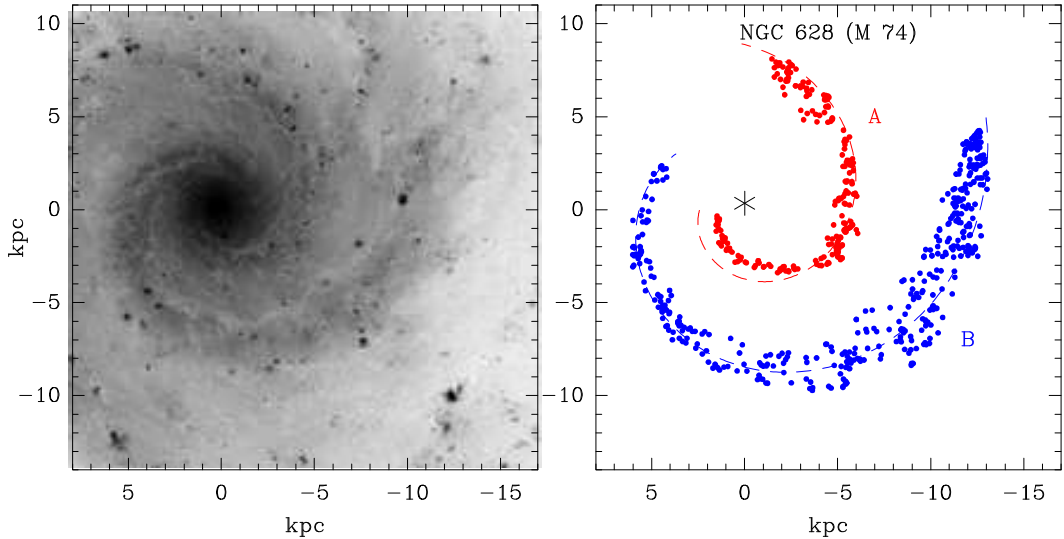}
\caption{\small
{\it Left panel:} Image of \Nste\ (\Msf) in an \Ha\ filter with North up and
East to the left.
{\it Right panel:} Locations of measured \HII\ regions. Dashed lines are a 
log-periodic spiral fitted to all \HII\ regions arms labeled A and B, separately.
        }
\label{fig:ngc628}
\end{figure}

The image of \Nste\ (\Msf)  is displayed in the left panel of Figure \ref{fig:ngc628}.
This galaxy has two prominent spiral arms that wind counterclockwise with increasing
distance from the galactic center.  We could trace 223 \HII\ regions in the arm 
labeled ``A'' in the figure and in Table \ref{table:ngc628_characteristics}
from a distance of about 2 to 8 kpc from the center.  Fitting the
log-periodic spiral form of Eq. (1) to all \HII\ region positions gives a 
global pitch angle of $-15\d4\pm0\d5$, although it is clear that a single pitch angle 
is not a good fit over the entire arm.  In the second arm, labeled ``B'', 
we located 427 \HII\ regions between distances of about 4 to 13 kpc and find a 
global pitch angle of $-14\d3\pm0\d2$.  This global fit is better than for arm A, but
still shows some small systematic deviations from a constant pitch angle form.

\begin{figure}[!htbp]
\epsscale{0.70} 
\plotone{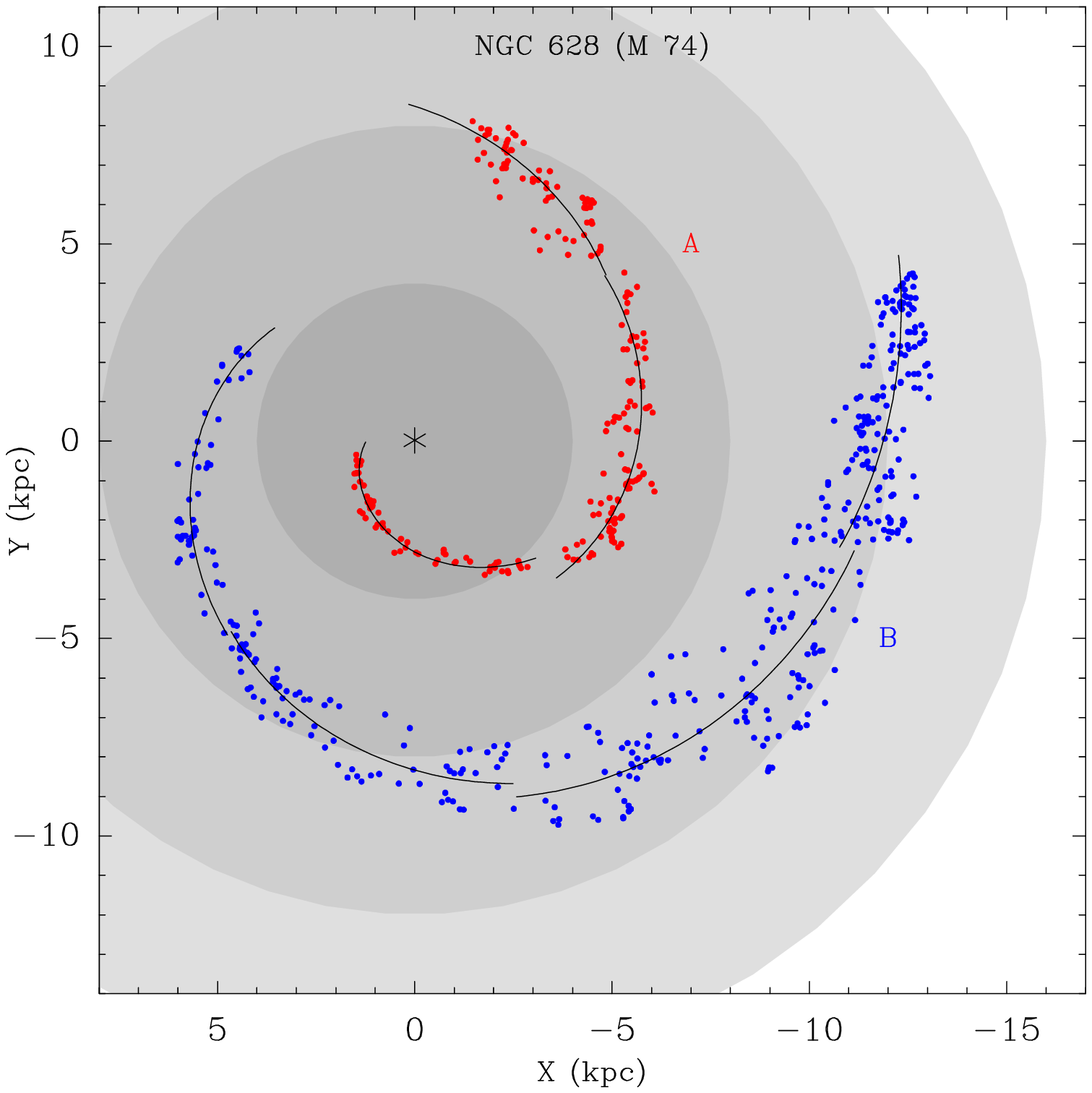}
\caption{\small
Locations of measured \HII\ regions for \Nste\ (\Msf) indicating
the segments of each arm fitted with a spiral.
Grey shaded annulae are spaced by 4 kpc and provide constant radius references.
        }
\label{fig:ngc628_detailed}
\end{figure}

We divided arm A into three segments and arm B into four segments, 
as indicated in Table \ref{table:ngc628_characteristics}.  Spiral fits
revealed significant changes in pitch angle among the segments in arm A,
but only small changes among segments in arm B.  For both arms, the estimated
arm widths increase with increasing radius, with the exception of
the last segment of arm B, where some narrowing occurs.  
These characteristics are evident from an inspection of Figure \ref{fig:ngc628_detailed}
where individual arm segment fits are displayed.

\begin{deluxetable}{crrrr}
\tablecolumns{5} \tablewidth{0pc}
\tablecaption{\Nste\ (\Msf) Spiral Arm Segment Characteristics}
\tablehead {
\colhead{Arm}  &\colhead{Azimuth Range} & \colhead{Mean Radius}& \colhead{Pitch Angle} & \colhead{Width}
\\
\colhead{}     &\colhead{(deg)}         &\colhead{(kpc)}       & \colhead{(deg)}       & \colhead{(kpc)} 
          }
\startdata
A             &$ 90\rightarrow225$       &$2.58\pm0.02$         &$-27.6\pm0.6$          &$0.14\pm0.02$ \\
...           &$225\rightarrow310$       &$5.60\pm0.04$         &$-9.4\pm1.0$           &$0.34\pm0.03$ \\
...           &$310\rightarrow360$       &$7.31\pm0.06$         &$-18.1\pm2.3$          &$0.42\pm0.04$ \\
\\
B             &$ 50\rightarrow125$       &$5.75\pm0.04$         &$-15.2\pm1.0$          &$0.27\pm0.03$ \\
...           &$125\rightarrow195$       &$7.57\pm0.05$         &$-15.9\pm1.1$          &$0.46\pm0.04$ \\
...           &$195\rightarrow255$       &$10.38\pm0.08$        &$-11.0\pm1.6$          &$0.87\pm0.06$ \\
...           &$255\rightarrow290$       &$12.02\pm0.05$        &$-15.4\pm1.2$          &$0.59\pm0.04$ \\
\enddata
\tablecomments{\footnotesize
Arm segments are labeled with letters A and B and are defined by the indicated azimuth range.
Azimuth increases east of north (counter clockwise from vertical in 
Figs. \ref{fig:ngc628} and \ref{fig:ngc628_detailed}.
                }
\label{table:ngc628_characteristics}
\end{deluxetable}

\subsection{\Nottt}

\begin{figure}[!htbp]
\epsscale{0.85} 
\plotone{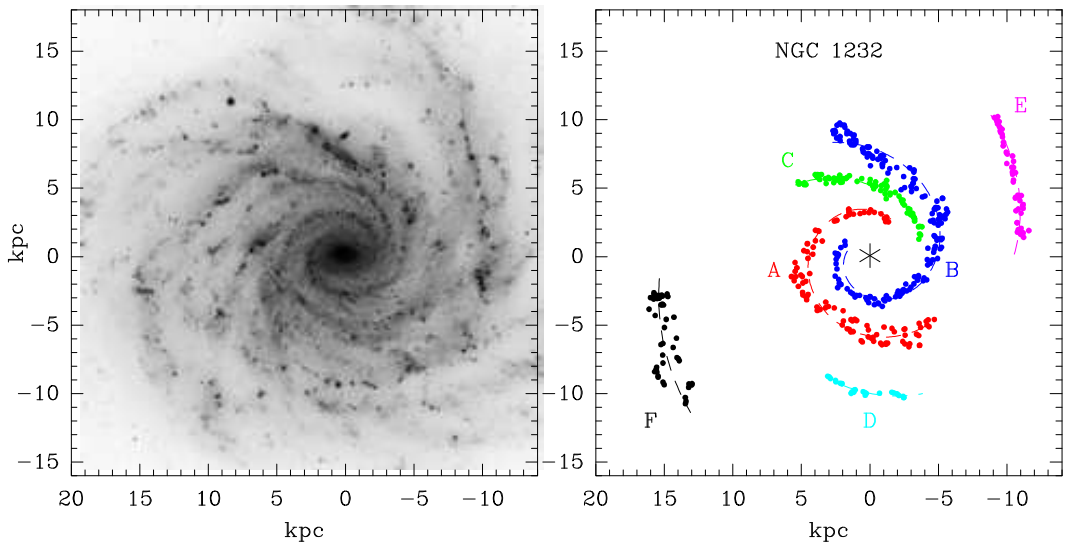}
\caption{\small
{\it Left panel:} Image of \Nottt\ in blue filter.
{\it Right panel:} Locations of measured \HII\ regions.  Dashed lines are a 
log-periodic spiral fitted to all \HII\ regions in arm segments A through F, separately.
        }
\label{fig:ngc1232}
\end{figure}

The left panel of Figure \ref{fig:ngc1232} shows the image of \Nottt, which displays
several spiral arms starting from near the galactic center and winding counterclockwise 
with increasing radial distance.  Some of these arms branch and/or connect with
arm segments at large distances.   We traced 105, 165, and 73 \HII\ regions in arms 
labeled ``A'' through ``C'', respectively, and 25, 50, and 48 \HII\ regions in arm 
segments ``D'' through ``F'', as labeled in the figure and in 
Table \ref{table:ngc1232_characteristics}. 
Arm segments E and F may connect to spiral arms A and C, although other possibilities exist.
Fitting a spiral form to all \HII\ regions in each arm (segment) gave global pitch
angles of $-9\d7\pm0\d5$, $-17\d3\pm0\d4$, $-18\d4\pm0\d5$, $-10\d9\pm0\d9$, $-16\d7\pm0\d6$, 
and $-10\d8\pm1\d9$ for arms or segments A through F, respectively.   Not only is
a single pitch angle inappropriate for all arms, the longer arms clearly show  
systematic deviations from a constant pitch angle along their length.

\begin{figure}[!htbp]
\epsscale{0.70} 
\plotone{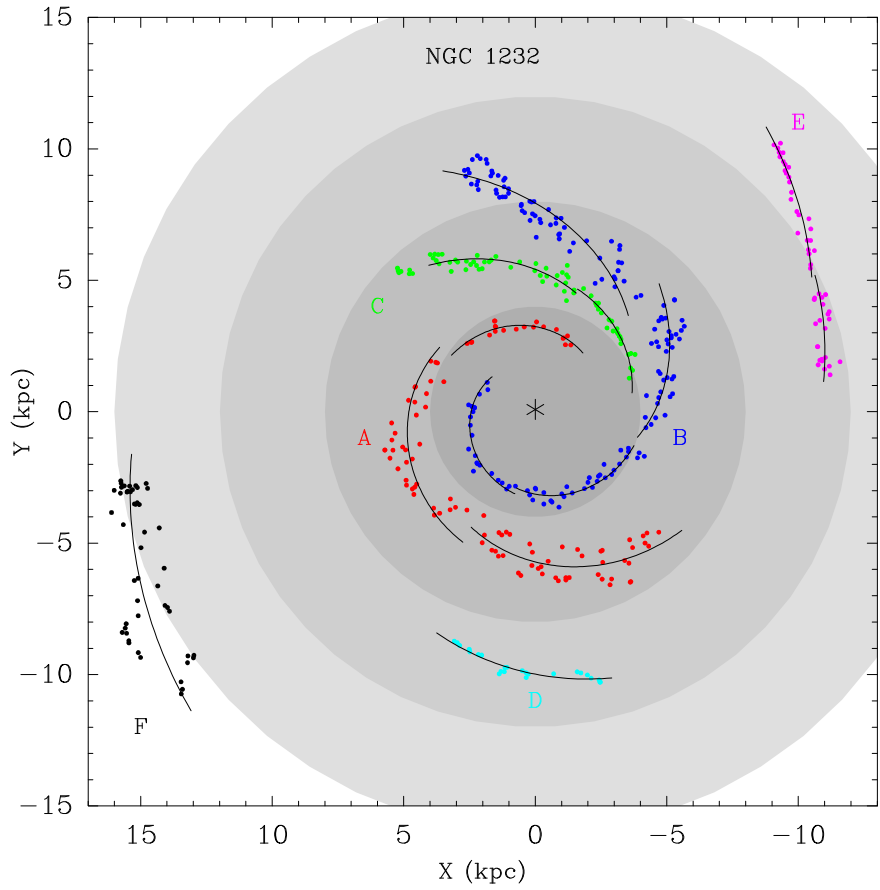}
\caption{\small
Locations of measured \HII\ regions for \Nottt\  indicating
the segments of each arm fitted with a spiral.
Grey shaded annulae are spaced by 4 kpc and provide constant radius references.
        }
\label{fig:ngc1232_detailed}
\end{figure}

We divided arms A, B, and C into two or more segments in order to estimate 
arm widths as a function of distance from the galactic center, as indicated in 
Table \ref{table:ngc1232_characteristics}.   The outer galaxy arm fragments
D and F were fit with one segment and E with two segments.  The spiral fits
revealed significant changes in pitch angle among segments within arms.
The widths for arms A, B, and C increase with distance from the galaxy center,
as is evident in Figure \ref{fig:ngc1232_detailed}.
Only the outer galaxy arm segment E shows evidence arm narrowing.

\begin{deluxetable}{crrrr}
\tablecolumns{5} \tablewidth{0pc}
\tablecaption{\Nottt\ Spiral Arm Segment Characteristics}
\tablehead {
\colhead{Arm}  &\colhead{Azimuth Range} & \colhead{Mean Radius}& \colhead{Pitch Angle} & \colhead{Width}
\\
\colhead{}     &\colhead{(deg)}         &\colhead{(kpc)}       & \colhead{(deg)}       & \colhead{(kpc)} 
          }
\startdata
A              &$-45\rightarrow\p55$     &$3.33\pm0.04$         &$\p-9.9\pm1.5$         &$0.16\pm0.03$ \\
...            &$ 45\rightarrow150$      &$5.01\pm0.08$         &$\p-8.9\pm2.3$         &$0.52\pm0.06$ \\
...            &$150\rightarrow230$      &$5.98\pm0.08$         &$-14.3\pm2.1$          &$0.51\pm0.06$ \\
\\
B              &$ 50\rightarrow165$      &$2.63\pm0.03$         &$-11.8\pm0.9$          &$0.10\pm0.02$ \\
...            &$165\rightarrow250$      &$3.49\pm0.04$         &$-10.8\pm1.6$          &$0.24\pm0.03$ \\
...            &$250\rightarrow315$      &$5.46\pm0.06$         &$-26.6\pm2.3$          &$0.33\pm0.04$ \\
...            &$312\rightarrow380$      &$7.64\pm0.07$         &$-30.0\pm1.5$          &$0.46\pm0.04$ \\
\\
C              &$-80\rightarrow-20$      &$4.38\pm0.03$         &$-14.7\pm1.3$          &$0.12\pm0.02$ \\
...            &$-20\rightarrow+35$      &$5.82\pm0.05$         &$-20.9\pm1.2$          &$0.27\pm0.03$ \\
\\
D              &$155\rightarrow195$      &$9.87\pm0.04$         &$-10.9\pm0.9$          &$0.16\pm0.03$ \\
\\
E              &$-85\rightarrow-65$      &$11.38\pm0.05$        &$-11.6\pm2.9$          &$0.24\pm0.04$ \\
...            &$-65\rightarrow-40$      &$12.83\pm0.03$        &$-22.0\pm1.0$          &$0.14\pm0.02$ \\
\\
F              &$ 95\rightarrow130$      &$16.23\pm0.10$        &$-10.8\pm1.9$          &$0.65\pm0.07$ \\
\enddata
\tablecomments{\footnotesize
Arm segments are labeled with letters A through F and are defined by the indicated azimuth range.
Azimuth is measured east of north (counter clockwise from vertical in 
Figs. \ref{fig:ngc1232} and \ref{fig:ngc1232_detailed}. Mean radius is the fitted galactocentric
distance at the average azimuth of the \HII\ regions in the arm segment.
                }
\label{table:ngc1232_characteristics}
\end{deluxetable}

\subsection{\Ntoef}

\begin{figure}[!htbp]
\epsscale{0.85} 
\plotone{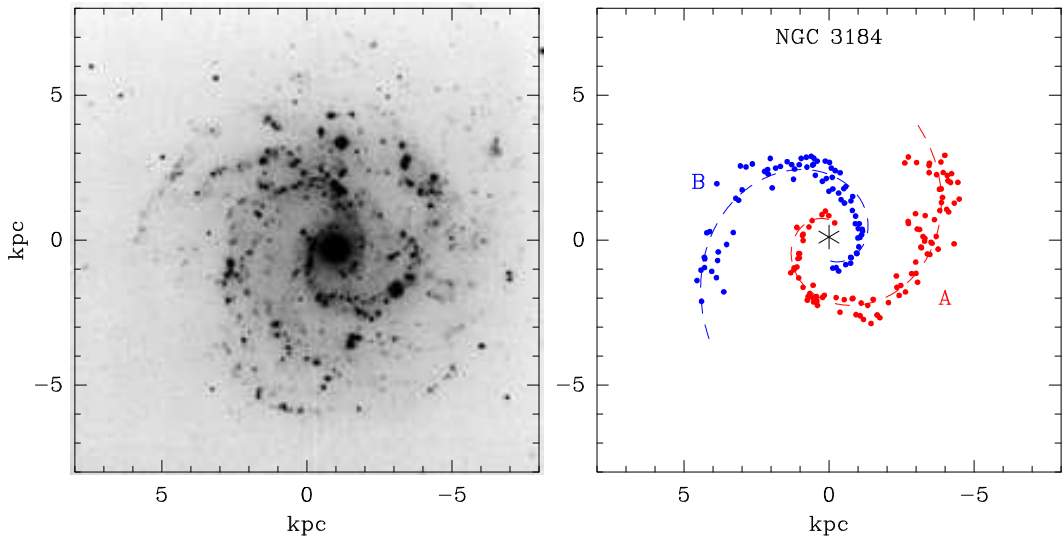}
\caption{\small
{\it Left panel:} Image of \Ntoef\ in an \Ha\ filter.
{\it Right panel:} Locations of measured \HII\ regions.  Dashed lines are a 
log-periodic spiral fitted to all \HII\ regions arms labeled A and B, separately.
        }
\label{fig:ngc3184}
\end{figure}

An image of \Ntoef\ is displayed in the left panel of Figure \ref{fig:ngc3184}.
This galaxy has two prominent and symmetrical spiral arms that wind counterclockwise 
with increasing distance from the galactic center.  We could trace 100 and 80 \HII\ regions in 
arms labeled ``A'' and ``B'', respectively, in the figure and in 
Table \ref{table:ngc3184_characteristics}, with distances of about 1 to 5 kpc from the center.
Fitting spirals to all \HII\ region positions for arms A and B separately gives global 
pitch angles of $-19\d6\pm0\d6$ and $-20\d2\pm0\d8$.  These values are consistent with 
each other.

\begin{deluxetable}{crrrr}
\tablecolumns{5} \tablewidth{0pc}
\tablecaption{\Ntoef\ Spiral Arm Segment Characteristics}
\tablehead {
\colhead{Arm}  &\colhead{Azimuth Range} & \colhead{Mean Radius}& \colhead{Pitch Angle} & \colhead{Width}
\\
\colhead{}     &\colhead{(deg)}         &\colhead{(kpc)}       & \colhead{(deg)}       & \colhead{(kpc)} 
          }
\startdata
A             &$  0\rightarrow\m180$    &$1.43\pm0.04$         &$-22.8\pm1.9$          &$0.19\pm0.03$ \\
...           &$180\rightarrow\m260$    &$2.79\pm0.07$         &$-14.9\pm3.7$          &$0.32\pm0.05$ \\
...           &$260\rightarrow\m320$    &$3.91\pm0.08$         &$-22.7\pm3.9$          &$0.47\pm0.05$ \\
\\
B             &$-180\rightarrow-130$    &$1.09\pm0.02$         &$ -5.4\pm1.2$          &$0.07\pm0.01$ \\
...           &$-30\rightarrow\m\p45$   &$2.57\pm0.05$         &$-26.9\pm2.5$          &$0.24\pm0.03$ \\
...           &$+45\rightarrow\m120$    &$3.95\pm0.09$         &$-12.2\pm3.3$          &$0.42\pm0.07$ \\
\enddata
\tablecomments{Arm segments are labeled with letters A and B and are defined by the indicated azimuth range.
Azimuth increases east of north (counter clockwise from vertical in 
Figs. \ref{fig:ngc3184} and \ref{fig:ngc3184_detailed}).
                }
\label{table:ngc3184_characteristics}
\end{deluxetable}

\begin{figure}[!htbp]
\epsscale{0.70} 
\plotone{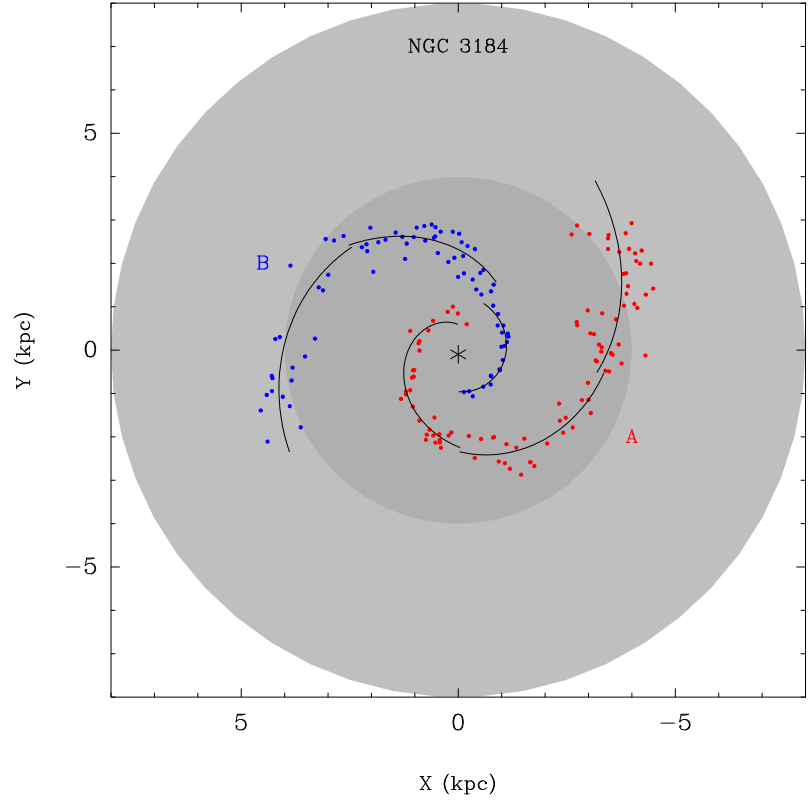}
\caption{\small
Locations of measured \HII\ regions for \Ntoef\ indicating
the segments of each arm fitted with a spiral.
Grey shaded annulae are spaced by 4 kpc and provide constant radius references.
        }
\label{fig:ngc3184_detailed}
\end{figure}

We divided arms A and B into three segments each, as indicated in 
Table \ref{table:ngc3184_characteristics}.    The spiral fits
revealed significant changes in pitch angle among segments within each arm.
The widths for both arms increase smoothly with distance from the galaxy center,
as is evident in Figure \ref{fig:ngc3184_detailed}.

\subsection{\Nfonf\  (\Mfo)}

\begin{figure}[!htbp]
\epsscale{0.85} 
\plotone{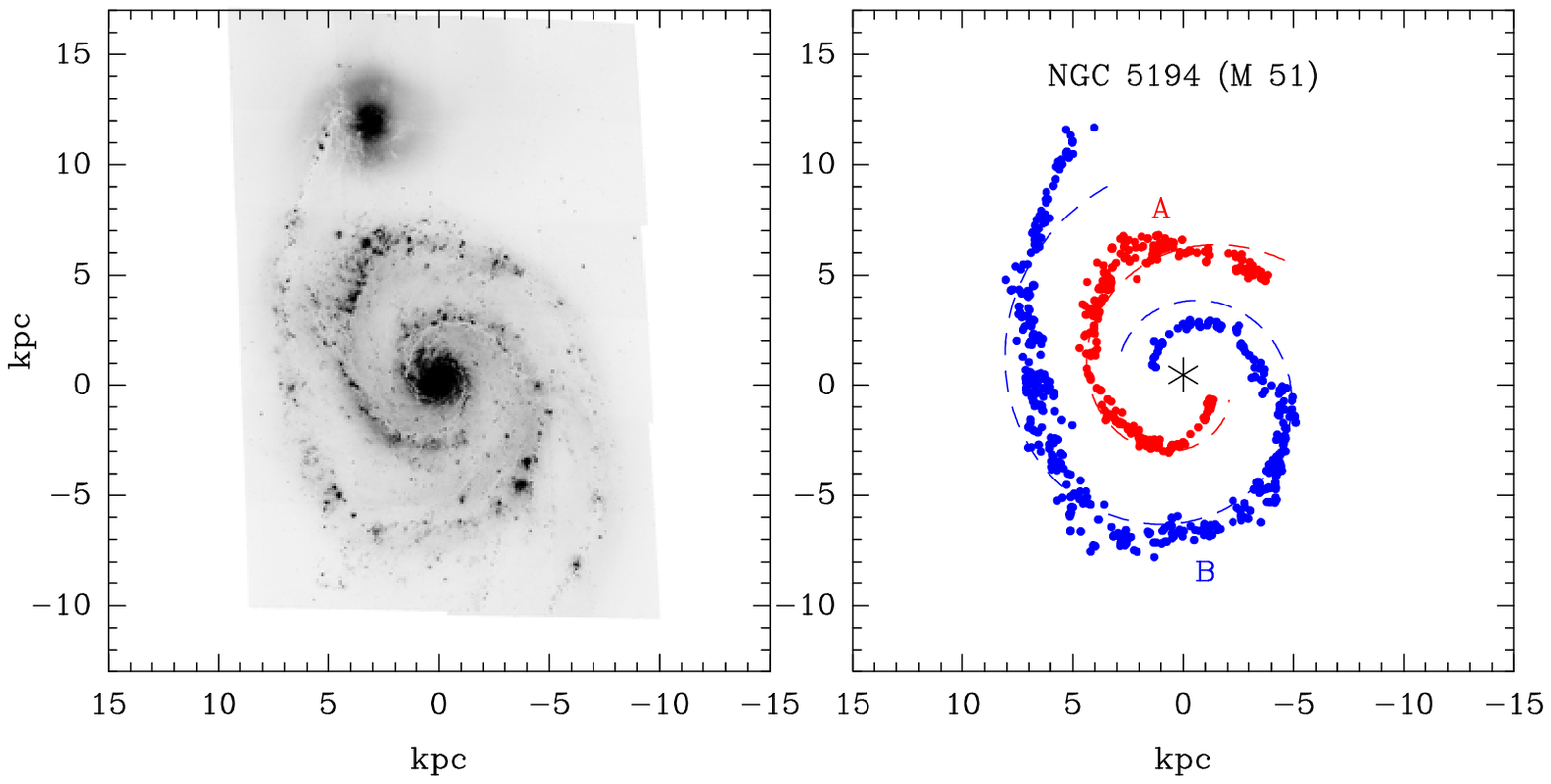}
\caption{\small
{\it Left panel:} Image of \Nfonf\ ((\Mfo) in \Ha\ filter.
{\it Right panel:} Locations of measured \HII\ regions.
        }
\label{fig:m51}
\end{figure}

The image of \Nfonf\ is displayed in the left panel of Figure \ref{fig:m51}.
This galaxy has two prominent spiral arms that wind clockwise 
with increasing distance from the galactic center.  We could trace 283 and 527 \HII\ regions in 
arms labeled ``A'' and ``B'', respectively, in the figure and in 
Table \ref{table:m51_characteristics}, 
with distances of about $2$ to $>10$ kpc from the center for arm B.   Fitting spirals to 
all \HII\ region positions for arms A and B separately gives global pitch angles of 
$+13\d4\pm0\d6$ and $+8\d3\pm0\d3,$ which are significantly different.  While systematic
deviations from the global model for arm A are modest, those for arm B are large.

\begin{deluxetable}{crrrr}
\tablecolumns{5} \tablewidth{0pc}
\tablecaption{\Nfonf\ (\Mfo) Spiral Arm Segment Characteristics}
\tablehead {
\colhead{Arm}  &\colhead{Azimuth Range} & \colhead{Mean Radius}& \colhead{Pitch Angle} & \colhead{Width}
\\
\colhead{}     &\colhead{(deg)}         &\colhead{(kpc)}       & \colhead{(deg)}       & \colhead{(kpc)} 
          }
\startdata
A             &$250\rightarrow\m170$      &$1.90\pm0.02$         &$+30.9\pm0.7$          &$0.07\pm0.01$ \\
...           &$165\rightarrow\m\p80$     &$3.29\pm0.02$         &$+15.0\pm0.9$          &$0.18\pm0.02$ \\
...           &$ 80\rightarrow\m\p30$     &$5.40\pm0.04$         &$+26.7\pm1.7$          &$0.26\pm0.02$ \\
...           &$ 30\rightarrow\o-15$      &$6.45\pm0.04$         &$ -9.1\pm2.0$          &$0.31\pm0.03$ \\
...           &$-15\rightarrow\o-40$      &$6.08\pm0.04$         &$ -4.0\pm4.7$          &$0.23\pm0.03$ \\
\\
B             &$ 65\rightarrow\o-40$      &$2.50\pm0.03$         &$+22.5\pm1.1$          &$0.14\pm0.02$ \\
...           &$-40\rightarrow\o-90$      &$3.42\pm0.04$         &$ +1.2\pm2.6$          &$0.20\pm0.03$ \\
...           &$-90\rightarrow-150$       &$5.39\pm0.03$         &$+19.4\pm1.0$          &$0.28\pm0.02$ \\
...           &$-150\rightarrow-200$      &$6.70\pm0.05$         &$ +8.6\pm1.6$          &$0.37\pm0.03$ \\
...           &$-200\rightarrow-250$      &$7.11\pm0.05$         &$-10.3\pm1.4$          &$0.43\pm0.04$ \\
...           &$-250\rightarrow-305$      &$7.05\pm0.04$         &$+19.2\pm1.3$          &$0.46\pm0.03$ \\
...           &$-305\rightarrow-345$      &$9.72\pm0.05$         &$+28.9\pm2.0$          &$0.22\pm0.03$ \\
\enddata
\tablecomments{Arm segments are labeled with letters A and B and are defined by the indicated azimuth range.
Azimuth increases east of north (counter clockwise from vertical in 
Figs. \ref{fig:m51} and \ref{fig:m51_detailed}).
                }
\label{table:m51_characteristics}
\end{deluxetable}

\begin{figure}[!htbp]
\epsscale{0.70} 
\plotone{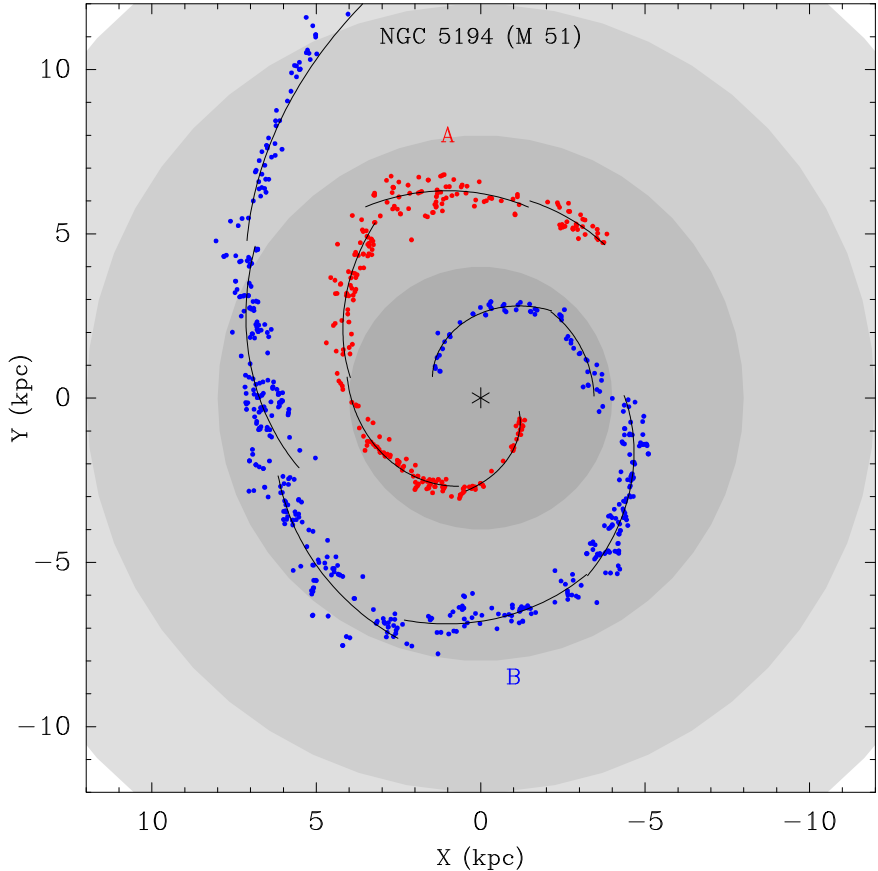}
\caption{\small
Locations of measured \HII\ regions for \Nfonf\ (\Mfo) indicating
the segments of each arm fitted with a spiral.  
Grey shaded annulae are spaced by 4 kpc and provide constant radius references.
        }
\label{fig:m51_detailed}
\end{figure}

We divided arms A and B into five and six segments, respectively, as indicated in 
Table \ref{table:m51_characteristics}.    The spiral fits
revealed significant changes in pitch angle among segments with each arm.
Overall, the widths for both arms increase smoothly with distance from the galaxy center,
as is evident in Figure \ref{fig:m51_detailed}.  However, both show a narrowing
in their last segment.

\section{Characteristics of Spiral Arms} \label{sect:characteristics}

Many studies present the properties of spiral arm pitch angles averaged over an entire 
galaxy \citep[e.g.,][]{Kennicutt:81,Savchenko:13,Davis:14}.   
Often these use Fourier or other automated techniques that can efficiently analyze 
large samples of galaxies using integrated light.  Our work differs by directly 
fitted spirals to the locations of \HII\ regions, which allows a detailed and 
robust evaluation of the characteristics of spiral arms without assuming
uniform properties over an entire arm or galaxy. 

\subsection {Pitch Angles}

\begin{figure}[!htbp]
\epsscale{0.65} 
\plotone{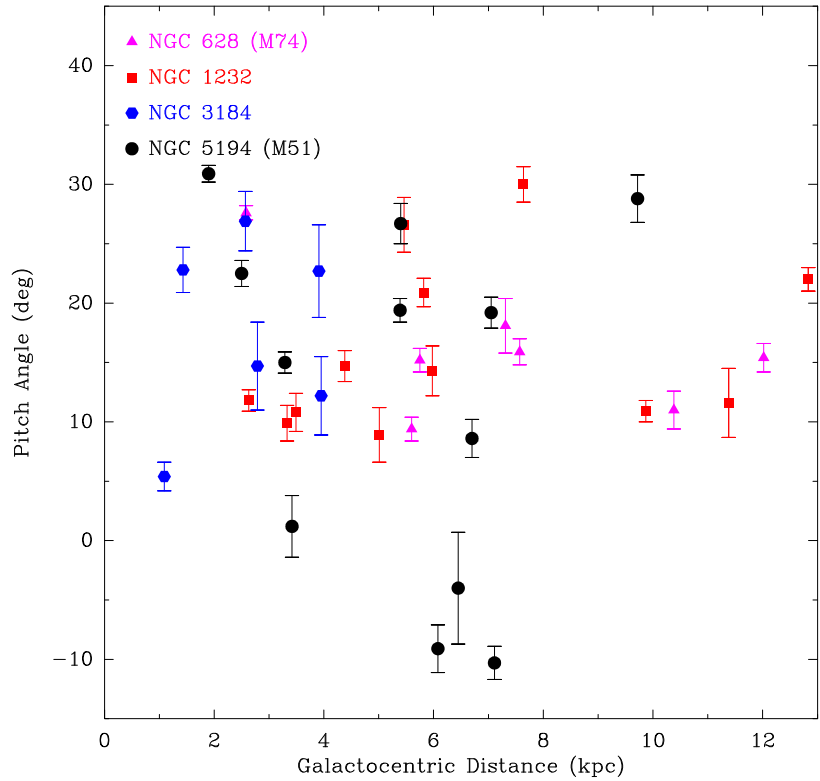}
\caption{\small
Pitch angles of spiral arms segments versus galactocentric distance for four 
galaxies measured in this work.  
Galaxy identifiers and plotting symbols are indicated in the upper left of the figure.
For the three galaxies that wind counter-clockwise
outward from the galactic center, we have plotted the negative of the measured
values.  There is a large scatter in pitch angles both among arms in different galaxies,
among arms within a galaxy, and even along a single spiral arm.
        }
\label{fig:pitchangles}
\end{figure}

Figure \ref{fig:pitchangles} shows the measured pitch angles of the segments of
spiral arms in our sample of galaxies.  Most pitch angles for these galaxies
are between about 10\deg\ and 30\deg, comparable to the range found by 
\citet{Kennicutt:81} for Sc-type galaxies.   Note that we also find large 
variations of pitch angles among arms within a spiral galaxy, as
well as along individual arms, confirming previous studies 
\citep[e.g.,][]{Russell:92,Ma:01,Savchenko:13,Davis:14}.

Interestingly, we find no evidence for a general change in pitch angle with 
galactocentric distance.  The $N$-body simulations of \citet{Grand:13} 
suggest that the pitch angle of an arm segment decreases with
the local rate of shear in a galaxy's rotation, and that shear rate generally
increases with radius.  Thus, one might expect a small trend of pitch angle
decreasing with galactocentric radius.  However, they also find that arm segments 
are short-lived and change pitch angles by $\sim10\deg$ over time scales
of $\sim0.1$ Gyr.  This may explain the apparent random variation in 
pitch angles within each galaxy in our sample, since different arm segments
would likely be observed at different ages.

The characteristics of pitch angles of Milky Way spiral arms are now
being revealed by trigonometric parallax measurements of high-mass star forming
regions from observations with the Very Long Baseline Array, the Japanese VERA
project, and the European VLBI Network.  Using parallax results from these arrays
yielded pitch angle estimates for sections of the following Milky Way arms:
Scutum arm: $\psi=19\d8\pm3\d1$ \citep{Sato:14}, 
Sagittarius arm: $\psi=7\d3\pm1\d5$ \citep{Wu:14},
Local arm: $\psi=10\d1\pm2\d7$ \citep{Xu:13},
Perseus arm: $\psi=9\d7\pm1\d5$ \citep{Zhang:13,Choi:14}, and
Outer arm: $\psi=14\d9\pm2\d7$ \citep{Hachisuka:14}.
The variation of pitch angle among segments of Milky Way spiral arms is
qualitatively similar to those of the four late-type spirals in this study. 
Currently the parallax data for the Milky Way typically
trace only arm segments $\sim5$ to $10$ kpc in length, which corresponds to
the scale over which we find relatively constant pitch angles in the external
galaxies.  However, as larger regions of the Milky Way are mapped, one might
expect to see pitch angle variations along its spiral arms.

\subsection {Arm Widths}

\begin{figure}[!htbp]
\epsscale{0.65} 
\plotone{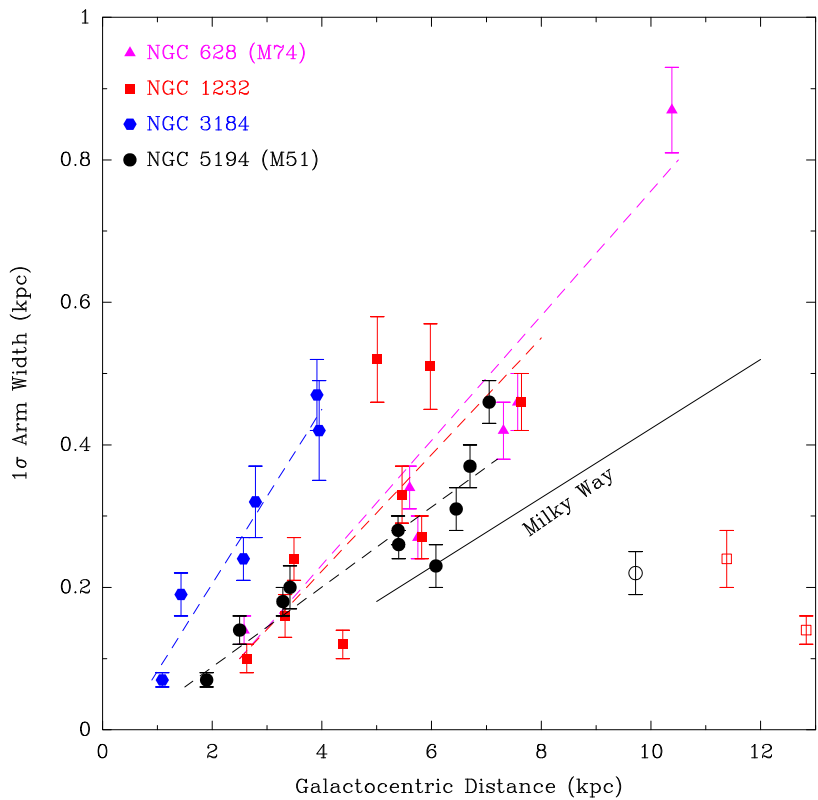}
\caption{\small
Widths of spiral arms versus galactocentric distance for four galaxies. 
Galaxy identifiers and plotting symbols are indicated in the upper left of the figure.
Only arms with two or more fitted segments are plotted. 
Open symbols indicate outermost arm distances if narrowing is observed.  
Dashed lines indicate trends for each galaxy.
The solid line is arm-width versus radius for the Milky Way, based on 
trigonometric parallax distances for high-mass star forming regions \citep{Reid:14}.
        }
\label{fig:widths}
\end{figure}

Figure \ref{fig:widths} shows the widths of arm segments (defined as the scatter 
perpendicular to the arm of their \HII\ regions) for the four 
galaxies in our sample.  Only arms with two or more segments are plotted. 
Except for the three outliers in the lower-right portion
of the figure, which indicate narrowing at the ends of some arms 
(to be discussed below), both individually and as a group these data indicate 
that spiral arms increase in width with distance from the center of their galaxy.
Clear examples of this phenomenon are seen in Figure \ref{fig:ngc628_detailed} for
arm B of \Nste\ (\Msf) and in Figure \ref{fig:ngc3184_detailed} for arm A of \Ntoef.
\citet{Lynds:70} reported a similar result for the widths of primary dust lanes
associated with spiral arms.

Unlike spiral pitch angles that vary significantly and randomly across a galaxy,
the widths of arm segments display a systematic variation with radius.
Thus, the observed arm-width versus radius trends should provide a clear observable 
to test and better understand $N$-body simulations.
At present we are unaware of arm width estimates from simulations.

A similar analysis by \citet{Reid:14} for Milky Way spiral arms noted that the scatter 
in \HII\ regions, based on trigonometric parallax distances, increased with
distance from the Galactic center.  This result motivated us to examine 
external spiral galaxies to determine if the Milky Way is typical or 
unusual in this respect.  Although based on this relatively small sample of external
galaxies, we conclude that spiral arms widths in the Milky Way are qualitatively 
similar to those in other spiral galaxies. 
 
Interestingly, in \Nste\ (\Msf) arm B, \Nottt\ arm E, and \Nfonf\ (\Mfo) arms A and B, 
we find that the trend of arm width increasing with distance reverses near the
outer tips of these arms.  These last arm segments are all significantly narrower 
than their immediately interior arm segment.  This narrowing appears related to
massive star formation dying out at large galactocentric radii.   In classical
spiral density-wave theory, this narrowing could be attributed to reaching
the radius of co-rotation of a spiral pattern with orbiting material.  If so,
\Mfo's arm A co-rotates at a radius of $\approx6$ kpc while arm B co-rotates
at a radius of $\approx9$ kpc.   The existence of (at least) two co-rotation radii
in \Mfo\ would argue against a single (global) pattern speed, 
consistent with the findings of \citet{Meidt:08}.

\subsection {Spiral Arm Formation}

The trend of spiral arm width increasing with galactocentric distance demonstrated
in this paper may provide a diagnostic characteristic for constraining models
of the origin of spiral arms.  The swing amplification mechanism \citep{Goldreich:65}
may lead to such a trend, as swinging inherently fans material outwards \citep{Toomre:81}. 
Whether or not global spiral density-wave theories naturally lead to the observed
growth of arm width has not been addressed.

In the galaxies we studied, a small number of spiral arms display nearly continuous 
arms with modest changes in pitch angle.  However, more commonly, we find the arms 
are much less regular.   Examination of Figures 2, 4, 6, and 8 reveals
that individual arms often break up into segments of $\sim5$ kpc length, and
different segments join to form bends, kinks, and quasi-linear sections  
and abrupt changes in pitch angle.   Similar conclusions have been reached based
on a variety of other observables (\eg \citealt{Russell:92,Waller:97,Chernin:99}).  
To these characteristics we now add variation in arm width.  These observations suggest 
that spiral arms are not purely global structures. 

Numerical simulations of galactic disks also show spiral characteristics that 
indicate non-global origin (\eg \citealt{Foyle:11,Grand:12}). 
In particular, the high-resolution $N$-body simulations by \cite{DOnghia:13} 
show that spiral structures can form in response to perturbations from 
giant molecular clouds in the disk, which then swing amplify to form arm segments, 
grow non-linearly, and then connect to produce long spiral-like patterns.  
Given the arm morphologies we have noted in late-type galaxies, 
such an origin for spiral arms seems an attractive mechanism.

\vskip 0.5truein
We thank the referee and E. D'Onghia for suggestions to improve the paper.

\end{document}